\documentclass[conference]{IEEEtran}
\ifCLASSINFOpdf
\else
\fi
\usepackage{algorithm}
\usepackage{algorithmic}
\usepackage{amsmath}
\usepackage{amssymb}
\usepackage{graphicx}
\usepackage[subfigure]{graphfig}
\usepackage{subfigure}
\usepackage{multirow}
\usepackage{float}
\usepackage{booktabs}
\usepackage{mathrsfs}
\usepackage[numbers,sort&compress]{natbib}
\makeatletter

\newcommand{\Rmnum}[1]{\expandafter\@slowromancap\romannumeral #1@}
\makeatother
\makeatletter
\makeatletter
\def\hlinew#1{%
	\noalign{\ifnum0=`}\fi\hrule \@height #1 \futurelet
	\reserved@a\@xhline}
\makeatother  
\begin{document}
	%
	\title{Improved Fuzzy $H_{\infty}$ Filter Design Method for Nonlinear Systems with Time-Varing Delay}


%
	
	
	%
	\author{\IEEEauthorblockN{Qianqian Ma\IEEEauthorrefmark{1},
	Li Li\IEEEauthorrefmark{2},
	Junhui Shen\IEEEauthorrefmark{1}, 
	Haowei Guan\IEEEauthorrefmark{3},
		Guangcheng Ma\IEEEauthorrefmark{1}, and
		Hongwei Xia\IEEEauthorrefmark{1}
}
	\IEEEauthorblockA{\IEEEauthorrefmark{1}School of Astronautics\\
Harbin Institute of Technology,
Harbin 150001, P. R. China\\ Email: maqq222008@hit. edu.cn}
	\IEEEauthorblockA{\IEEEauthorrefmark{2}School of Information Science and Engineering\\Harbin Institute of Technology at Weihai, Weihai 264200, P. R. China\\
	Email: lili406@hitwh.edu.cn}
	\IEEEauthorblockA{\IEEEauthorrefmark{3}Shanghai Institute of Satellite Engineering, Shanghai 201109\\
Email:	 davyfeng@sina.com}
}


	\maketitle
	


	%
	\IEEEpeerreviewmaketitle
	\begin{abstract}
		This paper investigates the fuzzy $H_{\infty}$ filter design issue for nonlinear systems with time-varying delay. In order to obtain less conservative fuzzy $H_{\infty}$ filter design method, a novel integral inequality is employed to replace the conventional Lebniz-Newton formula to analyze the stability conditions of the filtering error system. Besides, the information of the membership functions is introduced in the criterion to further relax the derived results. The proposed delay dependent filter design method is presented as LMI-based conditions, and corresponding definite expressions of fuzzy $H_{\infty}$ filter are given as well. Finally, a simulation example is provided to prove the effectiveness and  superiority of the designed fuzzy $H_{\infty}$ filter.
	\end{abstract}

	\section{Introduction}
	Filtering is playing a critical role in signal processing, and during last decades, a variety of filters have been developed, like Kalman filter \cite{steady1989}, $H_2$ filter \cite{Delay2003,sadat2019}, and 
	$H_{\infty}$ filter \cite{Y,JH}. Among them, $H_{\infty}$ filter has attracted considerable attention from researchers, since it can deal with systems with uncertainty, and it has no specific requirement for external noises.

	Moreover, time-delay often appears in practical systems, which can destroy the system performance and even cause instability of the control system. Therefore, the study of time-delay systems is of great importance \cite{H2008,Y,JH}. When dealing with analysis and synthesis problems of nonlinear systems with time-delay, the T-S fuzzy model\cite{short,An2015,ma2022,kong2021} is often employed, which can represent the nonlinear system as the weighted sum of some local linear models. Motivated by the parallel distribution compensation (PDC) methodology \cite{approach1996}, we assume that the fuzzy $H_{\infty}$ filter to be designed and the T-S fuzzy model have the same premise membership functions and the same number of rules in this paper. Thus, the stability analysis and synthesis can be facilitated.
	
	As the fuzzy $H_{\infty}$ filter has to guarantee the whole filtering system is asymptotically stable, the Lyapunov stability theory \cite{star} is applied, which will make the derived results conservative. To reduce conservatism, researchers have developed various methods \cite{H2008,Y,short,star,Wang2020,}. To mention a few, in \cite{H2008,Ma2017b}, a delay-dependent fuzzy $H_{\infty}$ filter design method was proposed for T-S fuzzy-model-based system with time-varying delay. However, in this paper, the Lyapunov-Krasovskii function candidate was chosen as a single Lyapunov function, to obtain more relaxed results, the literature  \cite{JH} adopted a fuzzy Lyapunov function to analyze the stability condition. In \cite{Y}, the fuzzy $H_{\infty}$ filter design approach was improved by estimating the upper bound of the derivative of Lyapunov function without ignoring any useful terms. On the basis of \cite{Y}, literature \cite{short} proposed a technique to obtain more accurate upper bound of the derivative of Lyapunov function.
	However, all the aforementioned literature used the inequalities derived from the Leibniz-Newton formula \cite{Y} to derive stability conditions, just like in some other literature \cite{further2007,new2009,Fuzzy2007, Qin2021,kong2019no,kong2019MIPR}. Though such methods can solve the fuzzy $H_{\infty}$ filter design problem, the derived results are conservative and there is little room left to further reduce the conservatism. Therefore, in this paper, we aim to propose a new less conservative fuzzy $H_{\infty}$ filter design method which is not based on the conventional Leibniz-Newton formula. Besides, to the best of our knowledge, no existing fuzzy $H_{\infty}$ filter design method has considered the information of membership functions, which motivate us to investigate a membership function dependent design method.
	
	To achieve our goal, the fuzzy $H_{\infty}$ filter to be designed and the T-S fuzzy model will be assumed to have the same premise membership functions and the same number of fuzzy rules. To reduce the conservatism of the design method, a novel integral inequality \cite{New2015} will be employed to replace the traditional Leibniz-Newton formula to deal with integral term $\int_{\alpha}^{\beta}\dot{x}^{T}(s)R\dot{x}(s)ds$ in stability analysis. Besides, the information of the membership functions will be taken into account in the derived criteria to further relax the derived results. 
	
	
	\section{Preliminaries}
	Consider a nonlinear system involving time-varying delay, which is described by the following $p$-rule T-S fuzzy model:
	\subsubsection*{Plant Rule $i$}
	IF $\psi_{1}(t) $ is $\mathcal{M}_{1}^i$ and
	$\psi_{2}(t)$ is $\mathcal{M}_{2}^i$ and $\ldots$ and $\psi_{m}(t)$ is $\mathcal{M}_{m}^i$, THEN
	\begin{eqnarray}
	\dot{x}(t)&=&A_{i}x(t)+A_{\tau{i}}x(t-\tau(t))+B_{i}w(t),\notag\\
	y(t)&=&C_ix(t) + C_{\tau{i}}x(t-\tau(t) + D_iw(t),\notag\\
		z(t)&=&E_ix(t)+E_{\tau{i}}x(t-\tau(t)), \notag
	\end{eqnarray}
	\begin{eqnarray}
	x(t)&=&\phi(t), \quad \forall  t\in{[-\tau_0,0]},  
	\end{eqnarray}
	where $i=1,2,\ldots,p$. $\psi_{\alpha}(t)(\alpha=1,2,\ldots,m)$ is the premise variable. $\mathcal{M}_{\alpha}^i$ is the fuzzy term of rule $i$ which corresponds to the function $\psi_{\alpha}$.  $m$ is a positive integer. And $ x(t)\in\mathbb{R}^{n} $ is the system state, $z(t)\in \mathbb{R}^{q}$ is the unknown signal to be estimated, $y(t)\in \mathbb{R}^{m}$ is the system output, $w(t)\in\mathbb{R}^{p}$ is the noise signal which is assumed to be arbitrary and satisfies $w(t)\in L_2\in[0,\infty)$. $A_i, A_{\tau{i}}, B_i,C_i, C_{\tau{i}}, D_i, E_i,E_{\tau{i}}$ are given system matrices. Time delay $ \tau(t) $ is a continuously differentiable function, satisfying the conditions:
	\begin{equation}\label{tau}
	0\leq \tau(t)<h, \qquad \dot{\tau}(t)\leq \rho.
	\end{equation}
	By fuzzy blending, the system dynamics can be presented as 	  	
	\begin{equation}\label{fuzzy model}
	\begin{aligned}
	\dot{x}(t)=&\sum_{i=1}^{p}\upsilon_{i}(\psi(t))[A_{i}x(t)+A_{\tau{i}}x(t-\tau(t))+B_{i}w(t)],\\
	y(t)=&\sum_{i=1}^{p}\upsilon_{i}(\psi(t))(C_ix(t) + C_{\tau{i}}x(t-\tau(t)+ D_iw(t)),\\
	z(t)=&\sum_{i=1}^{p}\upsilon_{i}(\psi(t))(E_ix(t)+E_{\tau{i}}x(t-\tau(t))).
	\end{aligned}
	\end{equation}
	where $\upsilon_{i}(\psi(t))=\Pi_{\alpha=1}^{m}\mu_{\mathcal{M}_{\alpha}^i}(\psi_{\alpha}(t))/\sum_{k=1}^{p}\Pi_{\alpha=1}^{m}\mu_{\mathcal{M}_{\alpha}^k}(\psi_{\alpha}(t))$ is the normalized membership function satisfying: $\sum_{i=1}^{p}\upsilon_{i}(\psi(t))=1$, $\upsilon_{i}(\psi(t))\geqslant0$; and $\mu_{\mathcal{M}_{\alpha}^i}(\psi_{\alpha}(\psi(t)))$ is the grade of membership function which corresponds to the fuzzy term $\mathcal{M}_{\alpha}^i$.

	
	Motivated by the parallel distribution compensation (PDC) methodology \cite{approach1996}, the fuzzy $H_{\infty}$ filter is assumed to have the same premise membership functions and the same number of fuzzy rules as the fuzzy model, which can be presented as:
	\subsubsection*{Filter Rule $j$} 
	IF $\psi_{1}(t) $ is $\mathcal{M}_{1}^j$ and
	$\psi_{2}(t)$ is $\mathcal{M}_{2}^j$ and $\ldots$ and $\psi_{m}(t)$ is $\mathcal{M}_{m}^j$, THEN
	
	\begin{equation}
	\begin{aligned}
	\dot{\hat{x}}(t)&=\hat{A}_{j}\hat{x}(t)+\hat{B}_{j}y(t),\\
	\hat{z}(t)&=\hat{C}_{j}\hat{x}(t),
	\end{aligned}
	\end{equation}	
	where $j=1,2,\ldots,p$. $\hat{x}(t)\in \mathbb{R}^{n}$ and $\hat{z}(t)\in \mathbb{R}^{q}$ are the state and output of the fuzzy $H_{\infty}$ filter respectively.  And $\hat{A}_{j},\hat{B}_{j},\hat{C}_{j}$ are the filter matrices of that will be designed.  
	
	Similarly, through fuzzy blending, the fuzzy $H_{\infty}$ filter to be designed can be presented as	
	\begin{equation}\label{fuzzyfilter}
	\begin{aligned}
	\dot{\hat{x}}(t)&=\sum_{j=1}^{p}\upsilon_{j}(\psi(t))(\hat{A}_{j}\hat{x}(t)+\hat{B}_{j}y(t)),\\
	\hat{z}(t)&=\sum_{j=1}^{p}\upsilon_{j}(\psi(t))\hat{C}_{j}\hat{x}(t).
	\end{aligned}
	\end{equation}
	
	
	
	According to (\ref{fuzzy model}) and (\ref{fuzzyfilter}), and define the augmented state vector as $\zeta(t)=[x^T(t),\hat{x}(t)]^T$ and $e(t)=z(t)-\hat{z}(t)$, we can obtain the $H_{\infty}$ filtering system as follows
	\begin{equation}\label{Closedfiltersystem}
	\begin{aligned}
	& \dot{\zeta}(t)=\bar{A}(t)\zeta(t)+\bar{A}_{\tau}(t)\zeta(t-\tau(t))+\bar{B}(t)w(t)), \\
	&   	e(t)=\bar{E}(t)\zeta(t) + \bar{E}_{\tau}(t)\zeta(t-\tau(t))),	
	\end{aligned}
	\end{equation}
	where $\zeta(0)=[\phi(t), \hat{x}_0]$ for $\forall t \in[-\tau_0,0]$, and
	\begin{flalign*}
	&\bar{A}(t)=\sum_{i=1}^{p}\sum_{j=1}^{p}\upsilon_{i}(\psi(t))\upsilon_{j}(\psi(t))\left[
	\begin{array}{cc}
	A_i & 0\\
	\hat{B}_jC_i & \hat{A}_j
	\end{array}
	\right]&
	\end{flalign*}
	\begin{flalign*}
	&\bar{A}_{\tau}(t)=\sum_{i=1}^{p}\sum_{j=1}^{p}\upsilon_{i}(\psi(t))\upsilon_{j}(\psi(t))\left[
	\begin{array}{cc}
	A_{\tau i} & 0\\
	\hat{B}_jC_{\tau i} & 0
	\end{array}
	\right]&\\
	&\bar{B}(t)=\sum_{i=1}^{p}\sum_{j=1}^{p}\upsilon_{i}(\psi(t))\upsilon_{j}(\psi(t))\left[
	\begin{array}{c}
	B_i\\   
	\hat{B}_jD_i
	\end{array}
	\right],\\
	&\bar{E}(t)=\sum_{i=1}^{p}\sum_{j=1}^{p}\upsilon_{i}(\psi(t))\upsilon_{j}(\psi(t))\begin{matrix}
	[E_i & -\hat{C}_j
	\end{matrix}], \\
	&\bar{E}_{\tau}(t)=\sum_{i=1}^{p}\sum_{j=1}^{p}\upsilon_{i}(\psi(t))\upsilon_{j}(\psi(t))[\begin{matrix}
	E_{\tau i} & 0
	\end{matrix}].
	\end{flalign*}
	
	Ergo, the fuzzy $H_{\infty}$ filter design problem that will be resolved in this paper can be summarized as follows:
	\subsubsection*{Fuzzy $H_{\infty}$  filter issue}   Design a fuzzy filter in the form of $(\ref{fuzzyfilter})$ satisfying the following two conditions: 
	
	(1) If $w(t)=0$, the filtering system  (\ref{Closedfiltersystem}) is asymptotically stable;
	
	(2) For a given scalar $\gamma >0$, if $\zeta(t)\equiv 0$ for $t\in[-h,0]$, the following $H_\infty$ performance can be  satisfied for all the $T>0$ and $w(t) \in L_2[0,\infty)$.
	\begin{eqnarray}\label{Hinf definition}
	\int_0^T\|e(t)\|^2dt\leq \gamma^2\int_0^T\|w(t)\|^2dt.
	\end{eqnarray}

	Besides, the following lemma is useful for the later deduction of the main results.
	\subsubsection*{Lemma 1 \cite{New2015}} It is assumed that $x$ is a differentiable function: $[\alpha,\beta]\rightarrow \mathbb{R}^{n}$. For $N_{1}, N_{2}, N_{3} \in \mathbb{R}^{4n\times n}$, and  $R \in \mathbb{R}^{n\times n }> 0$, the following inequality holds:
	\begin{equation}\label{lemma}
	-\int_{\alpha}^{\beta}\dot{x}^{T}(s)R\dot{x}(s)ds\leq \xi^{T} \Omega \xi,
	\end{equation}
	where
	\begin{align*}
	&\Omega = \tau(N_{1}R^{-1}N_{1}^{T}+\frac{1}{3}N_{2}R^{-1}N_{2}^{T}+\frac{1}{5}N_{3}R^{-1}N_{3}^{T})\\
	&+Sym\{N_{1}\Delta_{1}+N_{2}\Delta_{2}+N_{3}\Delta_{3}\},\quad\tau=\beta-\alpha, \\
	&e_{i}=\left[\
	\begin{matrix}
	0_{n\times(i-1)n}\quad I_{n} & 0_{n\times(4-i)n}
	\end{matrix}\right],\quad\Pi_{1}=e_{1}-e_{2},\\
	& \Pi_{2}=e_{1}+e_{2}-2e_{3},\quad\Pi_{3}=e_{1}-e_{2}-6e_{3}+6e_{4},\\
	&\xi=[\begin{matrix}
	x^{T}(\beta) & x^{T}(\alpha) & \frac{1}{d}\int_{\alpha}^{\beta}x^{T}(s)ds & \frac{2}{d^{2}}\int_{\alpha}^{\beta}\int_{\alpha}^{s}x^{T}(u)duds 
	\end{matrix}]^{T}.
	\end{align*} 	
	\section{Main Results}  	
	First, a sufficient stability condition for the filtering system (\ref{Closedfiltersystem}) will be derived.
	
	\subsubsection*{Lemma 2}
	Given constants $h$, $\rho$ and $\gamma>0$, the system (\ref{Closedfiltersystem}) is asymptotically stable with $w(t)\equiv0$, and satisfies the prescribed $H_{\infty}$ performance requirement (\ref{Hinf definition}), if there exist matrices $P=P^{T}\in\mathbb{R}^{2n\times 2n}$, $Y=Y^{T}\in\mathbb{R}^{2n\times 2n}$, $Z=Z^{T}\in\mathbb{R}^{2n\times 2n}$, such that the following inequality is feasible.
	\begin{equation}\label{lemma1}
	\Phi(t)=\left[\begin{matrix}
	\Xi(t) & \sqrt{h}\Gamma^{T}_{1}P & \Gamma^{T}_2(t)\\
	* & -PZ^{-1}P & 0\\
	* & * & -1
	\end{matrix}\right]<0,
	\end{equation}
	where
	\begin{flalign*}
	& \Xi(t)=\Lambda+\Xi_3(t),\\
	&\Lambda=\frac{3}{h}\left[
	\begin{matrix}
	-3Z & Z & 12Z & -10Z & 0\\
	* & -3Z & -8Z & 10Z & 0\\
	* & * & -64Z & 60Z & 0\\
	* & * & * & -60Z & 0\\
	* & * & * & * & 0
	\end{matrix}\right],&	\\
	&\Xi_{3}(t)=\left[\begin{matrix}
	P\bar{A}(t)+\bar{A}^{T}(t)P+Y & P\bar{A}_{\tau}(t) & 0 & 0 & P\bar{B}(t)\\
	* & -Y & 0 & 0 & 0\\
	* & * & 0 & 0 & 0\\
	* & * & * & 0 & 0\\
	* & * & * & * & -\gamma^{2}
	\end{matrix}\right],\\
	&\Gamma_1(t)=\left[\begin{matrix}
	\bar{A}(t) & \bar{A}_{\tau}(t) & 0 & 0 & \bar{B}(t)
	\end{matrix}\right],\\
	&\Gamma_2(t)=\left[\begin{matrix}
	\bar{E}(t) & \bar{E}_{\tau}(t) & 0 & 0
	\end{matrix}\right].
	\end{flalign*}
	\subsubsection*{Proof}
	Constructing the Lyapunov-Krasovskii function as follows:
	\begin{equation}\label{LKF}
	\begin{aligned}
	V(t)=&\zeta^{T}(t)P\zeta(t)+\int_{t-\tau(t)}^{t}\zeta^T(s)Y\zeta(s)ds\\
	&+\int_{-h}^{0}\int_{t+\theta}^{t}\dot{\zeta}(s)^T Z \dot{\zeta}(s)dsd\theta.	
	\end{aligned}
	\end{equation}  	
	Then the derivative of $V(t)$ can be obtained as
	\begin{equation}\label{LKFdot}
	\begin{aligned}
	\dot{V}(t)=& 2\zeta^{T}(t)P\dot{\zeta}(t)-(1-\dot{\tau}(t))\zeta^{T}(t-\tau(t))Y\zeta(t-\tau(t))\\
	&+h\dot{\zeta}(t)^TZ\dot{\zeta}(t)-\int_{t-h}^{t}\dot{\zeta}(s)^TZ\dot{\zeta}(s)ds.
	\end{aligned}
	\end{equation}  
	Applying Lemma 1 to the last term in the right hand side of (\ref{LKFdot}), we can obtain
	\begin{equation}
	\begin{aligned}
	&-\int_{t-h}^{t}\dot{\zeta}^{T}(s)Z\dot{\zeta}(s)ds <-\int_{t-\tau(t)}^{t}\dot{\zeta}^{T}(s)Z\dot{\zeta}(s)ds\\
	&\leq\mu^{T}(t)\left[\tau(t)M_{1}Z^{-1}M_{1}^{T}+\frac{\tau(t)}{3}M_{2}Z^{-1}M_{2}^{T}\right.&\\
	&+\frac{\tau(t)}{5}M_{3}Z^{-1}M_{3}^{T}+Sym\{M_{1}\Pi_{1}+M_{2}\Pi_{2}+M_{3}\Pi_{3}\}\bigg]\mu(t)&\\	
	&<\mu^{T}(t)\left[hM_{1}Z^{-1}M_{1}^{T}+\frac{h}{3}M_{2}Z^{-1}M_{2}^{T}+\frac{h}{5}M_{3}Z^{-1}M_{3}^{T}\right.&\\
	&+Sym\{M_{1}\Pi_{1}+M_{2}\Pi_{2}+M_{3}\Pi_{3}\}\bigg]\mu(t)&\\
	&=\mu^{T}(t)\left(\Xi_{1}+\Xi_{2}\right)\mu(t),&
	\end{aligned}
	\end{equation}
	where 
	\begin{flalign*}
	&\mu(t)=\\
	&[\begin{matrix}
	\zeta^{T}(t) & \zeta^{T}(t-\tau(t)) & \frac{1}{\tau(t)}\int_{t-\tau(t)}^{t}\zeta^{T}(s)ds & \theta_{1} & w(t)
	\end{matrix}]^{T},\\
	& \theta_1=\frac{2}{\tau^{2}(t)}\int_{t-\tau(t)}^{t}\int_{t-\tau(t)}^{s}\zeta^{T}(u)duds, \\	
	&\Xi_{1}=hM_{1}Z^{-1}M_{1}^{T}+\frac{h}{3}M_{2}Z^{-1}M_{2}^{T}+\frac{h}{5}M_{3}Z^{-1}M_{3}^{T},&\\
	&\Xi_{2}=Sym\{M_{1}\Pi_{1}+M_{2}\Pi_{2}+M_{3}\Pi_{3}\},&\\
	& e_{i}=\left[\
	\begin{matrix}
	0_{2n\times(i-1)2n}\quad I_{2n} & 0_{2n\times(5-i)2n}
	\end{matrix}\right].
	\end{flalign*}
	
	So according to (\ref{tau}) and  (\ref{Closedfiltersystem}), we can acquire
	\begin{equation}\label{ineV(t)}
	\begin{aligned}
	\dot{V}(t)&< 2\zeta^{T}(t)P[\bar{A}(t)\zeta(t)+\bar{A}_{\tau}(t)\zeta(t-\tau(t))+\bar{B}(t)w(t)]\\
	&\quad-(1-d)\zeta^{T}(t-\tau(t))Y\zeta(t-\tau(t))\\
	&\quad+h\dot{\zeta}(t)^TZ\dot{\zeta}(t)+\mu^{T}(t)(\Xi_{1}+\Xi_{2})\mu(t).
	\end{aligned}
	\end{equation}
	After some algebra, (\ref{ineV(t)}) can be expressed as the following compact form:
	\begin{equation}\label{deduction}
	\begin{aligned}
	&\dot{V}(t)+e^{T}(t)e(t)-\gamma^{2}w^{T}(t)w(t)<\mu^{T}(t)(\Xi_1+\Xi_2\\
	&+\Xi_3(t)+h\Gamma^{T}_1(t)Z\Gamma_1(t)+\Gamma^{T}_2(t)\Gamma_2(t))\mu(t),
	\end{aligned}
	\end{equation}
	where $\Xi_3(t)$, $\Gamma_1(t)$, $\Gamma_2(t)$ are defined in (\ref{lemma1}).
	
	So if
	\begin{equation}\label{transfer}
	\Xi_1+\Xi_2+\Xi_3(t)+h\Gamma^{T}_1Z\Gamma_1(t)+\Gamma^{T}_2\Gamma_2(t)<0,
	\end{equation}
	there will have
	\begin{equation}
	\dot{V}(t)+e^{T}(t)e(t)-\gamma^{2}w^{T}(t)w(t)< 0,
	\end{equation}
	which can be converted to
	\begin{equation} \label{Hperformance}
	\int_{0}^{L}(\Vert e(t)\Vert^{2}-\gamma^{2}\Vert w(t)\Vert^{2})dt+V(t)|_{t=L}-V(t)|_{t=0}\leq 0,
	\end{equation}
	as $V(t)|_{t=0}=0$ and $V(t)|_{t=L}\geq0$. So we have
	\begin{equation}
	\int_{0}^{L}\Vert e(t)\Vert^{2}dt\leq	\int_{0}^{L}\gamma^{2}\Vert w(t)\Vert^{2})dt,
	\end{equation}
	for all $L>0$, and any nonzero $w(t) \in L_2[0,\infty)$, which means the $H_{\infty}$ performance requirement is satisfied.
	
	Besides, to reduce computational complexity, we will eliminate free matrices by assuming
	\begin{equation}
	\begin{aligned}
	&M_1=\frac{1}{h}\left[
	-Z \quad Z \quad 0 \quad 0 \quad 0
	\right]^{T},\\
	&M_2=\frac{3}{h}\left[-Z \quad -Z \quad 2Z \quad 0 \quad 0
	\right]^{T},\\
	&M_3=\frac{5}{h}\left[
	-Z \quad Z \quad 6Z \quad -6Z \quad 0
	\right]^{T}.
	\end{aligned}
	\end{equation}	
	Then $\Xi_1+\Xi_2$ can be written as $\Lambda$, where $\Lambda$ is defined in (\ref{lemma1}). Applying the Schur Complement lemma, we can transfer (\ref{transfer}) to (\ref{lemma1}).
	
	Moreover, from inequalities (\ref{deduction}) and (\ref{transfer}), we can get $\dot{V}(t)<0$ when $w(t)\equiv0$, which means the filtering system (\ref{Closedfiltersystem}) is aymptotically stable. Thus, the proof of Lemma 2 is completed.
	
	\subsubsection*{Remark 1} It can be seen from the proof process that a novel integral inequality is applied to deal with the integral term $-\int_{t-h}^{t}\dot{\zeta}^{T}(s)Z\dot{\zeta}(s)ds$, which is tighter than other existing ones. Therefore, the derived results can be less conservative. Besides, the stability condition obtained has a simpler form, which means the proposed method can be more practical. Usually, the Leibniz-Newton formula \cite{further2007} is used to do this work, the introduction of the novel integral inequality provides another approach to further reduce conservatism, and improve the performance of the system.

	From the discussion above, we have got the sufficient condition for the existence of fuzzy $H_{\infty}$ filter. Next we will focus on  the fuzzy $H_{\infty}$ filter design for system (\ref{fuzzy model}).                                              
	\subsubsection*{Theorem 1}
	Given constants $h$, $\rho$, $\omega$ and $\gamma>0$, the system (\ref{Closedfiltersystem}) is asymptotically stable with $w(t)\equiv0$, and satisfies the prescribed $H_{\infty}$ performance requirement (\ref{Hinf definition}), if there exist matrices 
	\begin{equation}
	\tilde{P}=\tilde{P}^{T}=\left[\begin{matrix}
	P_{11} & \tilde{P}_{22}\\
	* & \tilde{P}_{22}
	\end{matrix}\right],
	\end{equation}
	$\tilde{Y}=\tilde{Y}^{T}\in\mathbb{R}^{2n\times 2n}$, $\tilde{Z}=\tilde{Z}^{T}\in\mathbb{R}^{2n\times 2n}$, such that the following LMIs $(\ref{theorem1})$ are feasible.
	\begin{equation}\label{theorem1}
	\Upsilon_{ij}+\Upsilon_{ji}<0,\quad i\leq j, \quad i,j=1,2,...,p,
	\end{equation}
	where
	\begin{flalign*}
	&\Upsilon_{ij}=\left[\begin{matrix}
	\tilde{\Xi}_{ij} & \sqrt{h}\tilde{\Gamma}^{T}_{1ij} & \tilde{\Gamma}^{T}_{2ij}\\
	* & -2\omega \tilde{P}+\omega^{2}\tilde{Z} & 0\\
	* & * & -1
	\end{matrix}\right],\\	
	& \tilde{\Xi}_{ij}=\tilde{\Lambda}+\tilde{\Xi}_{3ij},\\	
	&\tilde{\Lambda}=\frac{3}{h}\left[
	\begin{matrix}
	-3\tilde{Z} & \tilde{Z} & 12\tilde{Z} & -10\tilde{Z} & 0\\
	* & -3\tilde{Z} & -8\tilde{Z} & 10\tilde{Z} & 0\\
	* & * & -64\tilde{Z} & 60\tilde{Z} & 0\\
	* & * & * & -60\tilde{Z} & 0\\
	* & * & * & * & 0
	\end{matrix}\right],&\\	
	&\tilde{\Xi}_{3ij}=\left[\begin{matrix}
	Sym\{\lambda_{1ij}\}+\tilde{Y} & \lambda_{2ij} & 0 & 0 & \lambda_{3ij}\\
	* & -\tilde{Y} & 0 & 0 & 0\\
	* & * & 0 & 0 & 0\\
	* & * & * & 0 & 0\\
	* & * & * & * & -\gamma^{2}
	\end{matrix}\right],\\	
	&\tilde{\Gamma}_{1ij}=\left[\begin{matrix}
	\lambda_{1ij} & \lambda_{2ij} & 0 & 0 & \lambda_{3ij}
	\end{matrix}\right],\\
	&\tilde{\Gamma}_{2ij}=\left[
	\begin{matrix}
	\left[\begin{matrix}
	E_{i} & -\mathscr{C}_j
	\end{matrix}\right] & \left[\begin{matrix}
	E_{di} & 0
	\end{matrix}\right] & 0 & 0 & 0
	\end{matrix}\right],&\\
	&\lambda_{1ij}=\left[\begin{matrix}
	P_{11}A_i+\mathscr{B}_jC_i & \mathscr{A}_j\\
	\tilde{P}_{22}A_i+\mathscr{B}_jC_i & \mathscr{A}_j
	\end{matrix}\right],&\\
	&\lambda_{2ij}=\left[\begin{matrix}
	P_{11}A_{di}+\mathscr{B}_jC_{di} & 0\\
	\tilde{P}_{22}A_{di}+\mathscr{B}_jC_{di} & 0
	\end{matrix}\right],\lambda_{3ij}=\left[\begin{matrix}
	P_{11}B_i+\mathscr{B}_jD_i\\
	\tilde{P}_{22}B_i+\mathscr{B}_jD_i 
	\end{matrix}\right],&
	\end{flalign*}
	and in this case, the parameters of the fuzzy $H_{\infty}$ filter can be presented as
	\begin{equation}\label{parameter1}
	\hat{A'}_j=\tilde{P}^{-1}_{22}\mathscr{A}_j,\quad \hat{B'}_j=\tilde{P}^{-1}_{22}\mathscr{B}_j,\quad \hat{C'}_j=\mathscr{C}_j.
	\end{equation}
	\subsubsection*{Proof}
	
	For arbitrary constant $\omega$, the following inequality holds:
	\begin{equation}\label{omega}
	(\omega Z-P)Z^{-1}(\omega Z-P)\geq 0,
	\end{equation}	
	which can also be presented as
	\begin{equation}
	-PZ^{-1}P\leq-2\omega P+\omega^{2}Z.
	\end{equation}
	As a result, if the following inequality (\ref{lemma2up}) holds, the inequality (\ref{lemma1}) is true.
	\begin{equation}\label{lemma2up}
	\left[\begin{matrix}
	\Xi(t) & \sqrt{h}\Gamma^{T}_{1}P & \Gamma^{T}_2\\
	* & -2\omega P+\omega^{2}Z & 0\\
	* & * & -1
	\end{matrix}\right]<0.
	\end{equation}
	Then introduce a partition as
	\begin{equation}
	P=\left[
	\begin{array}{cc}
	P_{11}&P_{12}\\
	*&P_{22}
	\end{array}
	\right],
	\end{equation}
	where $P_{11}$, $P_{12}$ are assumed to satisfy: $P_{11}=P^{T}_{11}$, $P_{22}=P^{T}_{22}$, and $P_{12}$ is invertible by invoking small perturbation if it is necessary.
	
	Let
	\begin{equation}
	F=\left[
	\begin{matrix}
	I & 0\\
	* & P^{-T}_{22}P^{T}_{12}
	\end{matrix}\right],
	\end{equation}
	and $G=diag\{F,F,F,F,1\}$. Using $diag\{G,F,1\}$ and its transpose to post multiply and pre multiply (\ref{lemma2up}), then we can get
	\begin{equation}\label{lemma3}
	\begin{aligned}
	&\tilde{\Phi}(t)=\sum_{i=1}^{p}\sum_{j=1}^{p}\upsilon_{i}(\psi(t))\upsilon_{j}(\psi(t))\Upsilon_{ij}<0,
	\end{aligned}
	\end{equation}
	with the changes of variables as
	\begin{equation}\label{change}
	\begin{aligned}
	&\tilde{P}_{22}=P_{12}P^{-1}_{22}P^{T}_{12},\quad \tilde{P}=F^{T}PF=\left[\begin{matrix}
	P_{11} & \tilde{P}_{22}\\
	* & \tilde{P}_{22}
	\end{matrix}\right],&\\
	& \tilde{Y}=F^{T}YF,\quad \tilde{Z}=F^{T}ZF, 
	\quad\mathscr{A}_{j}=P_{12}\hat{A}_{j}P^{-T}_{22}P^{T}_{12},&\\
	& \mathscr{B}_j=P_{12}\hat{B}_j,\quad \mathscr{C}_j=\hat{C}_jP^{-T}_{22}P^{T}_{12},&
	\end{aligned}
	\end{equation}
	where $\Upsilon_{ij}$ is defined in (\ref{theorem1}).
	
	From (\ref{change}), we can obtain:
	\begin{equation}
	\begin{aligned}
	&\hat{A}_{j}=P_{12}^{-1}\mathscr{A}_jP^{-T}_{12}P^{T}_{22},\quad \hat{B}_j=P_{12}^{-1}\mathscr{B}_j,\\
	&\hat{C}_j=\mathscr{C}_jP_{12}^{-T}P_{22}^{T}.
	\end{aligned}
	\end{equation}
	As $\tilde{P}_{22}=P_{12}P^{-1}_{22}P^{T}_{12}$, through an equivalent transformation $P_{12}^{-T}P_{22}\hat{x}(t)$, we can obtain an admissible fuzzy $H_{\infty}$ realization as:
	\begin{equation}
	\begin{aligned}
	&\hat{A'}_{j}=P_{12}^{-T}P_{22}(P_{12}^{-1}\mathscr{A}_{j}P^{-T}_{12}P^{T}_{22})P_{22}^{-T}P^{T}_{12}=\tilde{P}^{-1}_{22}\mathscr{A}_{j},\\ &\hat{B'}_{j}=P_{12}^{-T}P_{22}(P_{12}^{-1}\mathscr{B}_{j})=\tilde{P}^{-1}_{22}\mathscr{B}_{j},\\
	& \hat{C'}_{j}=(\mathscr{C}_{j}P_{12}^{-T}P_{22}^{T})P_{22}^{-T}P^{T}_{12}=\mathscr{C}_{j}.
	\end{aligned}
	\end{equation}
	
	Besides, inequality (\ref{lemma3}) can also be denoted as
	\begin{equation}
	\begin{aligned}
	&\tilde{\Phi}(t)=\sum_{i=1}^{p}\sum_{j=1}^{p}\upsilon_{i}(\psi(t))\upsilon_{j}(\psi(t))\Upsilon_{ij}\\
	&=\sum_{i=1}^{p}\upsilon_{i}(\psi(t))^{2}\Upsilon_{ii}+\sum_{i=1}^{p}\sum_{i<j}^{p}\upsilon_{i}(\psi(t))\upsilon_{j}(\psi(t))(\Upsilon_{ij}+\Upsilon_{ji}),
	\end{aligned}
	\end{equation}
	where $\Upsilon_{ij}$ is defined in $(\ref{theorem1})$. 
	
	Therefore, if (\ref{theorem1}) holds, we can derive $	\tilde{\Phi}(t)<0,$
	which implies the filtering system (\ref{Closedfiltersystem}) is asymptotically stable, and it can satisfy the $H_{\infty}$ performance condition (\ref{Hinf definition}) as well. Thus, we finish the proof of Theorem 1.
	
	Theorem 1 has provided a feasible method to design fuzzy $H_{\infty}$ filter for system (\ref{fuzzy model}). However, as some inequality constraints is expanded in the deduction process, the derived result is conservative.
	To reduce conservatism of the filter design method, we will introduce the information of the  membership functions in the following criterion.
	
	\subsubsection*{Theorem 2}
	Given constants $h$, $\rho$, $\omega$ and $\gamma>0$, the system (\ref{Closedfiltersystem}) is asymptotically stable with $w(t)\equiv0$, and satisfies the prescribed $H_{\infty}$ performance requirement (\ref{Hinf definition}), if there exist matrices 
	
	\begin{equation}
	\tilde{P}=\tilde{P}^{T}=\left[\begin{matrix}
	P_{11} & \tilde{P}_{22}\\
	* & \tilde{P}_{22}
	\end{matrix}\right],
	\end{equation}
	$\tilde{Y}_i=\tilde{Y}^{T}_i\in\mathbb{R}^{2n\times 2n}$, $\tilde{Z}_i=\tilde{Z}^{T}_i\in\mathbb{R}^{2n\times 2n}$, $J_{ij}=J^{T}_{ij}\in\mathbb{R}^{(10n+2)\times (10n+2)}$, $K_{ij}=K^{T}_{ij}\in\mathbb{R}^{(10n+2)\times (10n+2)}$,  such that the following LMIs $(\ref{Theorem3})$ are feasible.
	\begin{equation}\label{Theorem3}
	\Omega_{ij}+\Omega_{ji}<0,\quad i\leq j, \quad i,j=1,2,...,p,
	\end{equation}
	where
	\begin{flalign*}
	&\Omega_{ij}=\Upsilon_{ij}-J_{ij}+K_{ij}+\sum_{a=1}^{p}\sum_{b=1}^{p}\bar{m}_{ab}J_{ab}-\sum_{k=1}^{p}\sum_{l=1}^{p}\bar{m}_{kl}K_{kl},&
	\end{flalign*}
	$\Upsilon_{ij}$ is defined in (\ref{theorem1}), and in this case, the parameters of the fuzzy $H_{\infty}$ filter can be expressed as
	\begin{equation}\label{parameter3}
	\hat{A'}_j=\tilde{P}^{-1}_{22}\mathscr{A}_j,\quad \hat{B'}_j=\tilde{P}^{-1}_{22}\mathscr{B}_j,\quad \hat{C'}_j=\mathscr{C}_j.
	\end{equation}
	\subsubsection*{Proof}
	In this part, we denote $\upsilon_{i}(\psi(t))\upsilon_{j}(\psi(t))$ as  $m_{ij}$, respectively to decrease the computational complexity. Besides, it is assumed that $\underline{m}_{ij}$ and $\bar{m}_{ij}$ represent the lower bound and upper bound of $m_{ij}$.
	
	From the discussion above, we have
	\begin{equation}
	\begin{aligned}
	&\dot{V}(t)+e^{T}(t)e(t)-\gamma^{2}w^{T}(t)w(t)<\mu^{T}(t)\tilde{\Phi}(t)\mu(t).
	\end{aligned}
	\end{equation}
	Through a straightforward computation, we can derive
	\begin{equation}
	\begin{aligned}
	&\mu^{T}(t)\tilde{\Phi}(t)\mu(t)=\sum_{i=1}^{p}\sum_{j=1}^{p}m_{ij}\mu^{T}(t)\Upsilon_{ij}\mu(t)\\
	&\leq \sum_{i=1}^{p}\sum_{j=1}^{p}m_{ij}\mu^T(t)\Upsilon_{ij}\mu(t) + \sum_{i=1}^{p}\sum_{j=1}^{p}(\bar{m}_{ij}\\
	&\quad-m_{ij})\mu^T(t)J_{ij}\mu(t)+\sum_{i=1}^{p}\sum_{j=1}^{p}(m_{ij}-\underline{m}_{ij})\mu^T(t)K_{ij}\mu(t)\\	
	&=\sum_{i=1}^{p}\sum_{j=1}^{p}m_{ij}\mu^{T}(t)(\Upsilon_{ij}-J_{ij}+K_{ij})\mu(t)\\
	&\quad+\sum_{i=1}^{p}\sum_{j=1}^{p}\bar{m}_{ij}\mu^{T}(t)J_{ij}\mu(t)-\sum_{i=1}^{p}\sum_{j=1}^{p}\underline{m}_{ij}\mu^{T}(t)K_{ij}\mu(t)\\ 
	&=\sum_{i=1}^{p}\sum_{j=1}^{p}m_{ij}\mu^{T}(t)(\Upsilon_{ij}-J_{ij}+K_{ij}+\sum_{a=1}^{p}\sum_{b=1}^{p}\bar{m}_{ab}J_{ab}\\
	&\quad-\sum_{k=1}^{p}\sum_{l=1}^{p}\bar{m}_{kl}K_{kl})\mu(t)=\sum_{i=1}^{p}\sum_{j=1}^{p}m_{ij}\mu^{T}(t)\Omega_{ij}\mu(t),
	\end{aligned}
	\end{equation}
	where $\Omega_{ij}$ is defined in $(\ref{Theorem3})$.
	
	Since
	\begin{equation}
	\begin{aligned}
	&\sum_{i=1}^{p}\sum_{j=1}^{p}m_{ij}\mu^{T}(t)\Omega_{ij}\mu(t)\\
	=&\sum_{i=1}^{p}m_{ii}\mu^{T}(t)\Omega_{ii}\mu(t)+\sum_{i=1}^{p}\sum_{i<j}^{p}m_{ij}\mu^{T}(t)(\Omega_{ij}+\Omega_{ji})\mu(t).
	\end{aligned}
	\end{equation}
	Consequently, if $(\ref{Theorem3})$ holds, we can derive that the following inequality holds
	\begin{equation}
	\begin{aligned}
	&\dot{V}(t)+e^{T}(t)e(t)-\gamma^{2}w^{T}(t)w(t)<0,
	\end{aligned}
	\end{equation}
	which means that the filtering system (\ref{Closedfiltersystem}) is asymptotically stable and the $H_{\infty}$ performance condition (\ref{Hinf definition}) can be satisfied, thus, we accomplish the proof of Theorem 2.
	
	\subsubsection*{Remark 2} In Theorem 2, the information of the membership functions is considered in the criterion, consequently, less conservative result can be obtained. However, compared with Theorem 1, the criterion presented in Theorem 2 is more complex, which implies that it will be more difficult to realize in engineering applications. Therefore, both Theorem 1 and Theorem 2 are meaningful.
	
	\section{Numerical Example}
	In this section, we will use one numerical example to illustrate the effectiveness of the proposed approach.
	\subsection{Example 1}
	Consider the example given in \cite{JH}, which can
	be presented as (1) with
	\begin{flalign*}
	& A_{1}=\left[\begin{matrix}
	-2.1 & 0.1\\ 1 & -2 
	\end{matrix}\right],
	A_{2}=\left[\
	\begin{matrix}
	-1.9 & 0\\ -0.2 & -1.1
	\end{matrix}\right],\\
	& A_{\tau 1}=\left[
	\begin{matrix}
	-1.1 & 0.1 \\ -0.8 & -0.9
	\end{matrix}\right], A_{\tau 2}=\left[
	\begin{matrix}
	-0.9 & 0 \\ -1.1 & -1.2
	\end{matrix}\right],B_{1}=\left[\
	\begin{matrix}
	1\\-0.2
	\end{matrix}\right],\\
	& B_{2}=\left[\
	\begin{matrix}
	0.3\\0.1
	\end{matrix}\right], C_{1}=\left[\
	\begin{matrix}
	1 & 0
	\end{matrix}\right],C_{2}=\left[\
	\begin{matrix}
	0.5 & -0.6
	\end{matrix}\right],\\
	& C_{\tau 1}=\left[\
	\begin{matrix}
	-0.8 & 0.6
	\end{matrix}\right],
	C_{\tau 2}=\left[\
	\begin{matrix}
	-0.2 & 1
	\end{matrix}\right], D_1=0.3,\\
	& D_2=-0.6, 
	E_{1}=\left[\
	\begin{matrix}
	1 & -0.5
	\end{matrix}\right],
	E_{2}=\left[\
	\begin{matrix}
	-0.2 & 0.3
	\end{matrix}\right],&\\
	&E_{\tau 1}=\left[\
	\begin{matrix}
	0.1 & 0
	\end{matrix}\right],
	E_{\tau 2}=\left[\
	\begin{matrix}
	0 & 0.2
	\end{matrix}\right].&
	\end{flalign*}
	and the membership functions are defined as
	\begin{flalign*}
	&\upsilon_{1}(x_{1}(t))=1-\frac{0.5}{1+e^{-3-x_{1}(t)}},\upsilon_{2}(x_{1}(t))=1-\omega_{1}(x_{1}(t)).&
	\end{flalign*} 
	
	Let $(\rho,\omega,h)=(0.2,2,0.5)$, using the criterion described in theorem 2, we can get the minimum attenuation level $\gamma=0.17$, and according to (\ref{Theorem3}) and (\ref{parameter3}), we can obtain a group of feasible filter parameters as follows
	\begin{equation*}
	\begin{aligned}
	& \hat{A}'_{1}=\left[\begin{matrix}
	-5.4988  &  0.7614\\
	0.6899 &  -1.6958
	\end{matrix}\right], \hat{A}'_{2}=\left[\begin{matrix}
	-0.0367 &  -8.6368\\
	-3.3984 & -10.8002
	\end{matrix}\right],\\
	& \hat{B}'_{1}=\left[\begin{matrix}
	-3.3721\\
	0.1692
	\end{matrix}\right],
	\hat{B}'_{2}=\left[\begin{matrix}
	-2.1468\\
	0.0148
	\end{matrix}\right],\\
	& \hat{C}'_{1}=\left[\begin{matrix}
	-1.0777 &   0.1460
	\end{matrix}\right], \hat{C}'_{ 2}=\left[\begin{matrix}
	-0.5006 &   0.0419
	\end{matrix}\right].
	\end{aligned}
	\end{equation*}
	
	Note that different $(h,\omega)$ can yield different value of minimum attenuation level $\gamma$, and to fully illustrate the superiority of the proposed method, we will use Theorem 2 in this paper and other recently developed methods to find minimum attenuation level $\gamma$. All the computational result are summarized in Table \Rmnum{1}-\Rmnum{3}.
	\begin{table}[!h]
		\caption{ The minimum attenuation level $\gamma$ for $\omega=2$}\label{tb1}
		\centering
		\begin{tabular}{c c c c c c c }
			\hlinew{0.75pt}
			method & $h=0.5$ & $h=0.6$ & $h=0.8$ & $h=1$ \\
			\hline  			
			\cite{JH} & 0.25 & 0.25 & 0.27 & 0.29\\
			\cite{short} & 0.24 & 0.25 & 0.25 & 0.26\\
			\cite{An2015} & 0.23 & 0.24 & 0.25 & 0.25\\			
			Th. 2 & 0.17 & 0.19 & 0.21 & 0.24\\
			\hlinew{0.75pt}
		\end{tabular}
	\end{table}  	
	
	\begin{table}[!h]
		\caption{ The minimum attenuation level $\gamma$ for $\omega=5$}\label{tb2}
		\centering
		\begin{tabular}{c c c c c c c }
			\hlinew{0.75pt}
			method & $h=0.5$ & $h=0.6$ & $h=0.8$ & $h=1$ \\
			\hline  			
			\cite{JH} & 0.24 & 0.24 & 0.25 & 0.26\\
			\cite{short} & 0.24 & 0.24 & 0.25 & 0.26\\
			\cite{An2015} & 0.23 & 0.24 & 0.24 & 0.25\\			
			Th. 2 & 0.18 & 0.20 & 0.21 & 0.22\\
			\hlinew{0.75pt}
		\end{tabular}
	\end{table}  
	
	\begin{table}[!h]
		\caption{ The minimum attenuation level $\gamma$ for $\omega=20$}\label{tb4}
		\centering
		\begin{tabular}{c c c c c c c }
			\hlinew{0.75pt}
			method & $h=0.5$ & $h=0.6$ & $h=0.8$ & $h=1$ \\
			\hline  			
			\cite{JH} & 0.26 & 0.28 & 0.44 & $--$\\
			\cite{short} & 0.25 & 0.26 & 0.35 & 0.45\\
			\cite{An2015} & 0.23 & 0.24 & 0.25 & 0.25\\			
			Th. 2 & 0.23 & 0.23 & 0.24 & 0.25\\
			\hlinew{0.75pt}
		\end{tabular}
	\end{table}   		  			
	where $--$ denotes that the minimum attenuation level $\gamma$ does not exist.  	
	
	From Table \Rmnum{1}-\Rmnum{3}, we can see that the designed approach in this paper can produce smaller value of minimum attenuation level $\gamma$ than those in \cite{Y,short,JH,H2008}, which implies the proposed method in this paper is less conservative than those in \cite{Y,short,JH,H2008}.
	
	\subsubsection*{Remark 3} There are two reasons for the less conservative results. First, a novel integral inequality (\ref{lemma}) is introduced, which is tighter than the conventional integral inequalities derived from the Leibniz-Newton formula, and less conservative stability conditions can be derived. Second, a membership functions dependent technique is used, which can help further relax the results.
	
	\section{Conclusions}
	This paper investigates the fuzzy $H_{\infty}$ filter design issue for nonlinear systems with time-varying delay. The T-S fuzzy model has been used to represent the dynamics of the nonlinear time-delay system. And a novel integral inequality which is tighter than the inequalities derived from the Leibniz-Newton formula has been applied in the deduction process. Motivated by the PDC methodology, the fuzzy filter in this paper has been allowed to have the same membership functions as the fuzzy model. Besides, the information about the membership functions has been introduced to reduce the conservatism. And in the last, a simulation example has been given to illustrate the effectiveness and the superiority of the proposed criteria.


\begin{thebibliography}{1}
		
		\bibitem{steady1989}
		D. S. Bernstein and W. M. Haddad, ``Steady-state Kalman filtering with
		an $H_{\infty}$ error bound," \emph{Systems \& Cotrol Letters}, vol. 12, pp. 9-16, 1989.
		
		\bibitem{Delay2003}
		H. Gao and C. Wang, ``Delay-dependent robust $H_{\infty}$ and
		$L_{2}$$-$$L_{\infty}$ filtering for a class of uncertain nonlinear time-delay systems," \emph{IEEE Transactions on Automatic Control}, vol. 48, no. 9, pp. 1661-1666, 2003.
		
		
		\bibitem{Y}
		Y. K. Su, B. Chen, C. Lin, and H. G. Zhang, ``A new fuzzy H$_\infty$ filter
		design for nonlinear continuous-time dynamic systems with time-varying
		delays," \emph{Fuzzy Sets and Systems}, vol. 160, pp. 3539-3549, 2009.
		
		\bibitem{JH}
		J. H. Zhang, Y. Q. Xiao, and R. Tao, ``New results on $H_{\infty}$ filtering for fuzzy time-delay systems," \emph{IEEE Transactions on Fuzzy System}, vol. 17, no. 1, pp. 128-137, 2009.
		
		\bibitem{H2008}
		C. Lin, Q. G. Wang, T. H. Lee and B. Chen, ``$H_{\infty}$ filter design for nonlinear systems with time-delay through T–S fuzzy model approach," \emph{IEEE Transactions on Fuzzy System}, vol. 16, no. 3, pp. 739-745, 2008.		
		
		\bibitem{short}
		S. J. Huang, X. Q. He and N. N. Zhang, ``New Results on $H_{\infty}$ Filter Design for Nonlinear Systems With time-delay via T–S Fuzzy Models," \emph{IEEE Transactions on Fuzzy Systems}, vol. 19, no. 1, pp. 193-199, 2011.		
		
		\bibitem{An2015}		
		T. Zhou and X. Q. He, ``An improved $H_{\infty}$ filter design for nonlinear systems
		described by T-S fuzzy models with
		time-varying delay," \emph{International Journal of Automation and Computing}, vol. 12, no. 6, pp. 671-678, 2015.			
		
		\bibitem{approach1996}
		H. O. Wang, K. Tanaka and M. F. Griffin, ``An approach to fuzzy control of nonlinear systems: stability and the design issues," \emph{IEEE Transactions on Fuzzy Systems}, vol. 4, no. 1, pp. 14-23, 1996.
		
		\bibitem{star}
		C. Lin, Q. G. Wang, T. H. Lee and B. Chen, ``$H_{\infty}$ Filter Design for Nonlinear Systems with Time-Delay Through T–S Fuzzy Model Approach," \emph{IEEE Transactions on Fuzzy Systems}, vol. 16, no. 13, pp. 739-746, 2008.		
		
		\bibitem{further2007}
		Y. He, Q. G. Wang, L. Xie and C. Lin, ``Further improvement of free-weighting matrices technique for systems with time-varying delay," \emph{IEEE Transactions on Automatic Control}, vol. 52, no. 2, pp. 293-299, 2007.	 		  		
		
		\bibitem{new2009}
		J. Qiu, F. Gang and J. Yang, ``A new design of delay-dependent robust $H_{\infty}$
		filtering for discrete-time T–S fuzzy systems
		with time-varying delay," \emph{International Journal on Fuzzy Systems}, vol. 17, no. 5, pp. 1044-1058, 2009.
		
		\bibitem{Fuzzy2007}
		C. Lin, Q. G. Wang, T. H. Lee and Y. He, ``Fuzzy weighting-dependent approach to $H_{\infty}$ filter
		design for time-delay fuzzy systems," \emph{IEEE Transactions on Signal Processing}, vol. 55, no. 6, 2007.
		
		\bibitem{New2015}
		B. Z. Hong, H. Yong, W. Min and S. Jinhua, ``New results on stability analysis for systems with discrete distributed delay," \emph{Automatica}, vol. 60, pp. 189-192, 2015.	 
		
		
		\bibitem{kong2019no}
		L. Kong, A. Ikusan, R. Dai, J. Zhu, and D. Ros, ``A no-reference image quality model for object detection on embedded cameras," \emph{International Journal of Multimedia Data Engineering and Management (IJMDEM)}, vol. 10, no. 1, pp. 22-39, 2019.	 
		
		\bibitem{kong2019MIPR}
		L. Kong, A. Ikusan, R. Dai, and J. Zhu, ``Blind image quality prediction for object detection," \emph{2019 IEEE Conference on Multimedia Information Processing and Retrieval (MIPR)},  pp. 216-221, 2019.	 

		\bibitem{sadat2019}
		M. Sadat, R. Dai, L. Kong, J. Zhu, ``QoE-aware multi-source video streaming in content centric networks," \emph{IEEE Transactions on Multimedia}, vol. 22, no. 9, pp. 2321--2330, 2019.

		\bibitem{kong2021}
	L. Kong, A. Ikusan, R. Dai, D. Ros, ``An image quality adjustment framework for object detection on embedded cameras," \emph{International Journal of Multimedia Data Engineering and Management (IJMDEM)}, vol. 12, no. 3, pp. 1-19, 2021.

   
  		\bibitem[Wang et~al.(2020)]{Wang2020}	
		L. Wang, B. Zong, Q. Ma, W. Cheng, J. Ni, W. Yu, Y. Liu, D. Song, H. Chen and Y. Fu.\newblock Inductive and unsupervised representation learning on graph structured object.\newblock  \emph{International conference on learning representations}, 2020.
		
		
  		\bibitem[Qin et~al.(2021)]{Qin2021}	
	C. Qin, L. Wang, Q. Ma, Y. Yin, H. Wang, and Y. Fu.\newblock Contradictory structure learning for semi-supervised domain adaptation.\newblock  \emph{Proceedings of the 2021 SIAM International Conference on Data Mining (SDM)}, 576-584, 2021.	
	
			\bibitem[Ma(2022)]{ma2022}	
		Q. Ma.\newblock First-order optimization methods for networked high dimensional systems.\newblock  \emph{Ph.D Thesis, Department of ECE, Boston University}, 2022.
		
		      		\bibitem[Ma et~al.(2017)]{Ma2017b}	
	Q. Ma, H. Xia, G. Ma, Y. Xia, and C. Wang.\newblock Improved stability and stabilization criteria for TS fuzzy systems with distributed time-delay.\newblock  \emph{International Conference on Data Mining and Big Data}, 517-526, 2017.		
		
	\end{thebibliography}
\end{document}